%% file: paper.tex
\pdfoutput=1

\documentclass{acm_proc_article-sp}
\newif\ifDemo
\newif\ifFull 
\newif\ifAnon
\newif\ifDraft
\Demofalse
\Fulltrue
\Draftfalse
\Anonfalse

\usepackage{amsmath}
\usepackage{amssymb}
\usepackage{graphicx}
\usepackage{algorithm}
\usepackage{algorithmic}
\usepackage{xspace}
\usepackage{url}
\usepackage{rotating}
\usepackage{multirow}
\usepackage{subfigure}

\newcommand{\name}{Haze}
\newcommand{\noises}{\mathcal{N}}
\newcommand{\perm}{\mathcal{P}}
\newcommand{\auths}{\mathcal{A}}
\newcommand{\encrypt}{\mathsf{encrypt}}
\newcommand{\decrypt}{\mathsf{decrypt}}
\newcommand{\msg}{m}
\newcommand{\true}{\mathsf{true}}

\newcommand{\pk}{\mathsf{PK}}
\newcommand{\sk}{\mathsf{SK}}

\newcommand{\cats}{\mathcal{C}}
\newtheorem{theorem}{Theorem}

\begin{document}

\title{\name: Privacy-Preserving Real-Time Traffic Statistics}

\ifAnon
\else
\numberofauthors{3}
\author{
%
\alignauthor
Joshua Brown
\alignauthor
Olga Ohrimenko
\alignauthor
Roberto Tamassia
\and
\{jwsbrown,olya,rt\}@cs.brown.edu\\
Brown University\\
Providence, RI 02912
}
\fi
\date{}

\maketitle
\begin{abstract}
\input{abstract}

\end{abstract}

\input{intro}

\input{model}
\ifFull
\input{crypto}
\fi
\input{protocol}
\ifFull
\input{security}
\fi
\input{prevwork}

\input{results}
\ifDraft

\fi

\ifFull
\input{acks}
\fi

\bibliographystyle{abbrv}
\bibliography{paper} 

\balancecolumns
\end{document}

%% file: abstract.tex
We consider \ifFull traffic-update \fi mobile applications that let users learn
traffic conditions based on reports from other users.
\ifFull
These applications are becoming increasingly
popular (e.g., Waze reported 30 million users in 2013) since they
aggregate real-time road traffic updates from actual
users traveling on the roads.
\fi
However, the providers of
these mobile services have access to such sensitive
information as timestamped locations and movements of its users.
In this paper, we \ifFull describe \fi \emph{\name}, a \ifFull protocol \fi for
traffic-update applications that supports the creation of traffic
statistics from user reports while protecting the privacy of the
users.
\ifFull
\name~relies on a small subset of users to jointly aggregate
encrypted speed and alert data and report the result to the service provider.
We use jury-voting protocols based on threshold cryptosystem
and differential privacy techniques
to hide user data from anyone participating in the protocol
while allowing only aggregate information to be extracted and sent to
the service provider.
We show that \name~is effective in practice by
developing a prototype implementation
and performing experiments on 
a real-world dataset of car trajectories.
\else
\fi

%% file: intro.tex
\section{Introduction}

\begin{figure}[t]
\begin{center}
\includegraphics[scale=0.20]{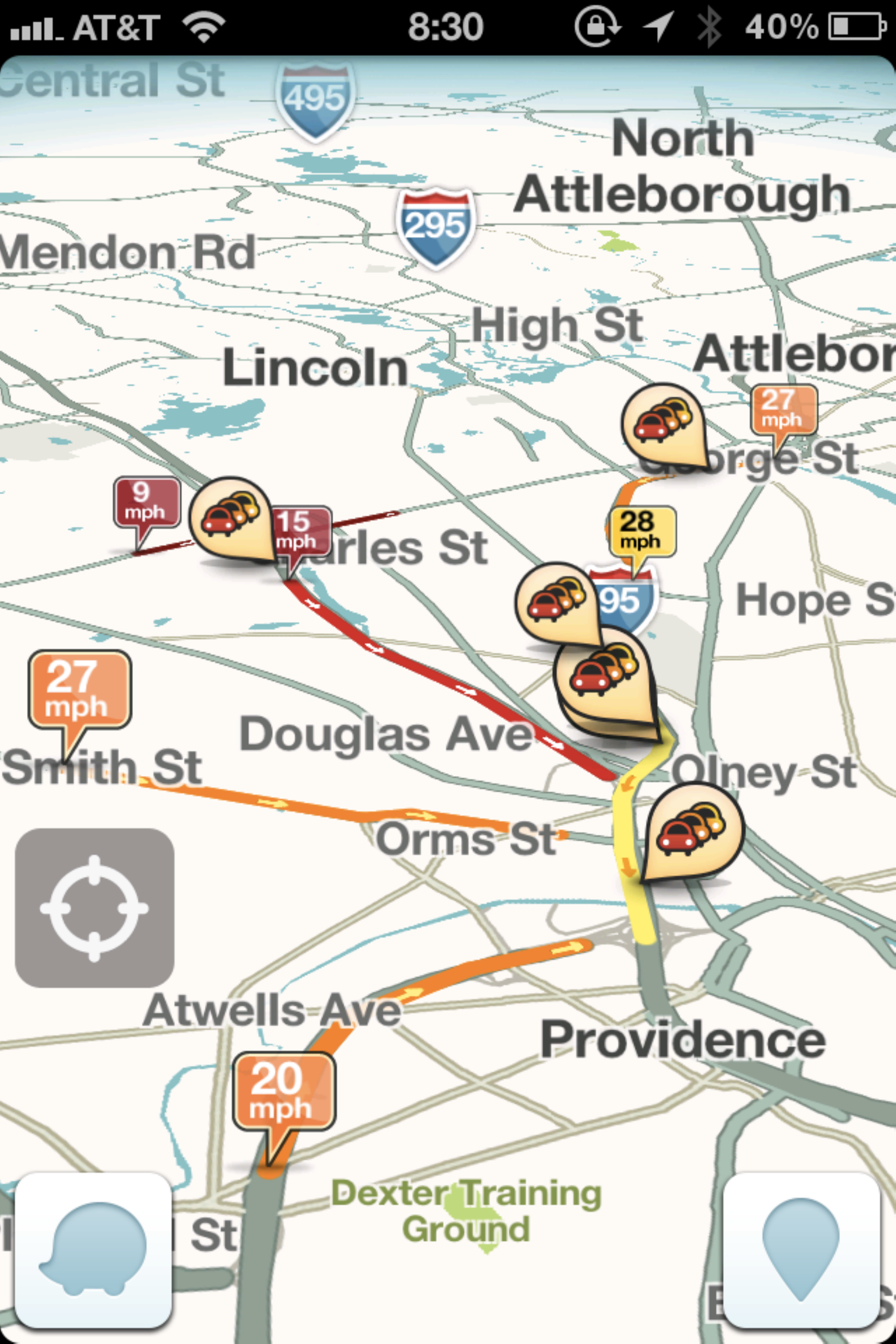}
\caption{Snapshot of Waze traffic update.
The icon with multiple cars shows traffic jam
and the color of the roads shows the traffic speed.}
\label{fig:waze}
\end{center}
\vspace{-14pt}
\end{figure}

\ifFull
The interest in having access to our location based data
has been confirmed one more time in June of this year
when media articles reported that Google outbid Facebook to acquire
navigation startup Waze~\cite{waze} for \$1 billion.%
\footnote{``Google Confirms Waze Maps App Purchase'' in Wall Street
  Journal, Jun 11 2013.}
Waze is a mobile application that provides
\else
\fi
 real-time traffic
data to its users. 
This data comes from the users
themselves, who contribute fresh data by uploading
their GPS coordinates.
\ifDemo
\else
Crowdsourcing navigation with 30 million users allows real time
updates and, hence, is very useful when picking a route
to avoid delays
\fi
 (see Figure~\ref{fig:waze}).

User location data, however, contains
very sensitive information.
Analyzing
GPS travel data can reveal the location of an individual's
house and work, since there is a limited number of possible routes
that one can take~\cite{ekxr-lgtd-13}.
Moreover, since these data points are timestamped,
one can also learn home departure and arrival  times,
as well as trip duration and purpose~\cite{k-inf-07,bkb-extr-12}.
\ifFull
In the case of Waze, user location data can be joined
with his Google account and potentially reveal
even more information.
Moreover, court cases of unreported tracking confirm
that mobile users are not comfortable
with sharing their location data~\cite{wired-13}.
\fi

\begin{figure}[th]
\begin{center}
\includegraphics[scale=0.35]{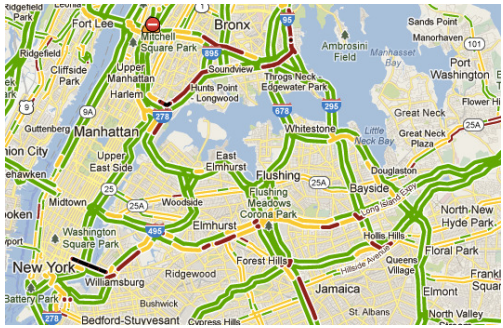}
\caption{Snapshot of Google Maps traffic update. As in Waze
the color shows the traffic speed:
high (green), medium (yellow) and low (red).
}
\label{fig:googlemaps}
\end{center}
\vspace{-18pt}
\end{figure}

The challenge in preserving user privacy in services such
as Waze is due to the nature of their functionality.
In order to give quality service
to its users, location data has to be collected for aggregation
and statistical processing.
However, we observe that precise user data is not required to provide
traffic information such as current speed on the roads or presence of congestions.
Consider the traffic information displayed by Waze and Google Maps 
in Figures~\ref{fig:waze} and~\ref{fig:googlemaps}, respectively.
It consists of color-coded road segments
representing traffic level, which is 
enough to plan routes effectively.
To report such statistics, the
service provider only needs to
know if enough drivers are traveling
at full speed, or if a significant number of users has reported
slight or severe delays.
Also, note that the maps display 
no data for the less-traveled roads.  Conveniently for this type of application, areas which 
do not receive enough observations to allow reliable statistics are typically not a source 
of travel delays and can be safely ignored. 

Motivated by the goal of providing privacy protection
in crowdsourced real-time traffic 
update services,
we propose a privacy-preserving version
of such services, called Haze.
Given that the data used by the navigation mobile app
is already contributed by the drivers,
we show that the aggregation functionality also can
be outsourced
and distributed among the users.
Moreover, we  develop a method that,
allows meaningful traffic statistics to
be extracted even when operating on
user encrypted data.

\begin{figure}[t]
\begin{center}
\includegraphics[scale=0.33]{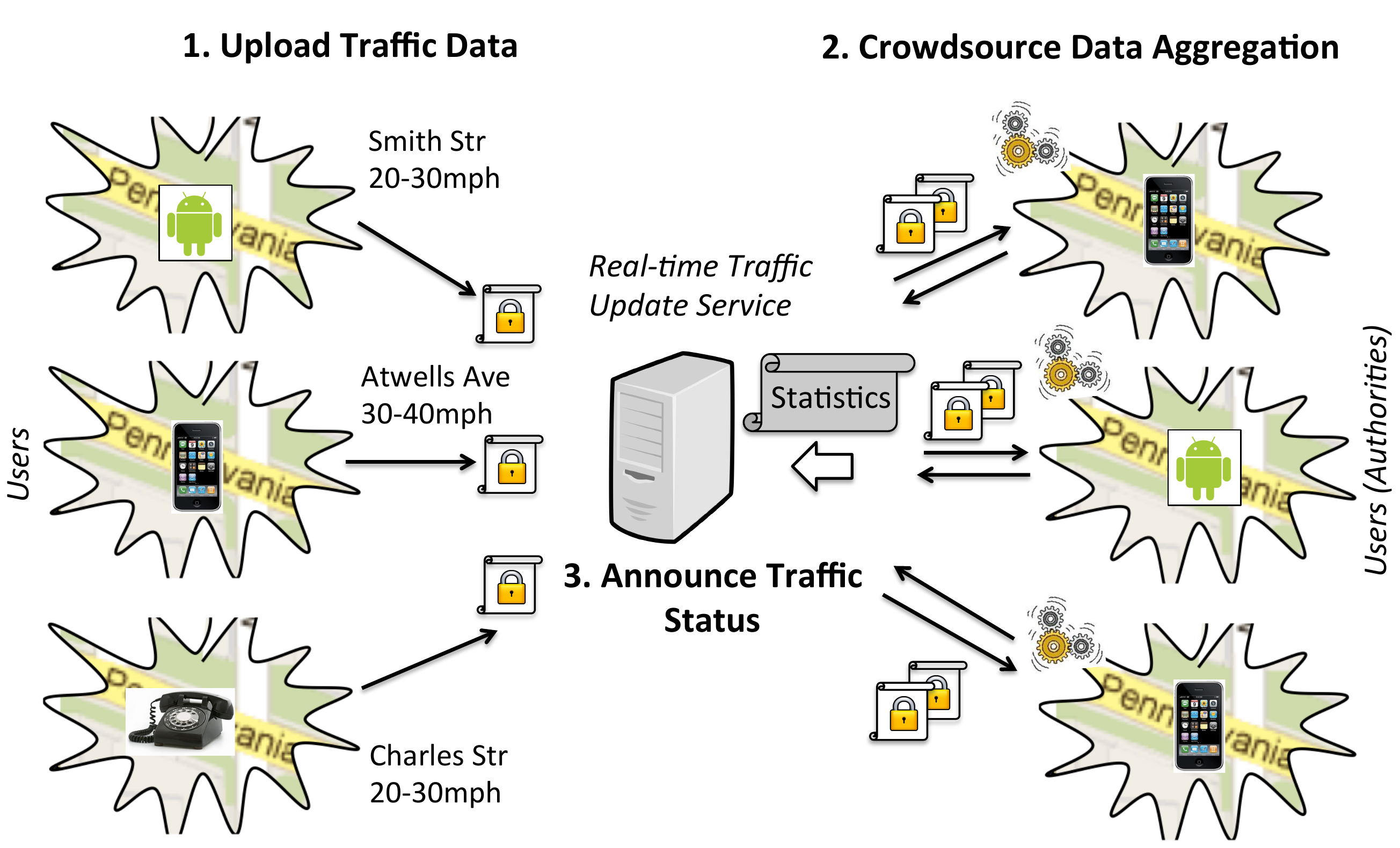}
\caption{Haze overview: 1.~Traffic data is
uploaded encrypted by the users.
2.~Other users (authorities) process encrypted data to extract
aggregate statistics.
3.~The traffic statistics are published by the service
provider. }
\label{fig:haze}
\end{center}
\vspace{-18pt}
\end{figure}

Haze is a history-independent protocol that is invoked
by the service provider whenever she
wishes to update the traffic map.
Figure~\ref{fig:haze} gives a high-level overview
of the Haze framework.
The task of traffic map update is outsourced to the users and is split in several phases:
some users upload their data (users on the left in Figure~\ref{fig:haze}),
while others aggregate this data (users, aka authorities, on the right),
and the final result is reported back to users by the service provider.
We reduce data upload to a voting protocol where instead
of sending the exact data value, the user casts a vote for
an observation that fits his data, e.g., instead of sending 45mph
as his current speed he votes for the range 40--50mph.
User vote is protected by encryption performed
by the user before the upload. Hence, neither the service
provider nor authorities have access to plain
data of other users.
Authorities tally the encrypted votes for all observations
and report those that are deemed significant, i.e.,
enough users have reported the same observation.

Haze also takes into account privacy leaks that data encryption
cannot solve, as well some of the problems it introduces.
First, a potential privacy leak due to multiple runs of Haze
arises
when a user is not present at all invocations,
e.g., because the user reached his destination.
The Haze protocol guarantees differential privacy~\cite{dmns-diff-06},
i.e., it protects the privacy of each individual user by
adding noise to the reported statistics so that
a user's observation cannot significantly change the traffic
status.
Secondly, submitting encrypted data may
encourage malicious users to report invalid data,
e.g., by submitting speed reports for multiple roads.
Haze prevents this behavior by enforcing users
to prove the validity of their submitted observations.
\ifFull
The dataflow of Haze is shown in Figure~\ref{fig:datalifecycle}.

\begin{figure}[t]
\begin{center}
\includegraphics[scale=0.5]{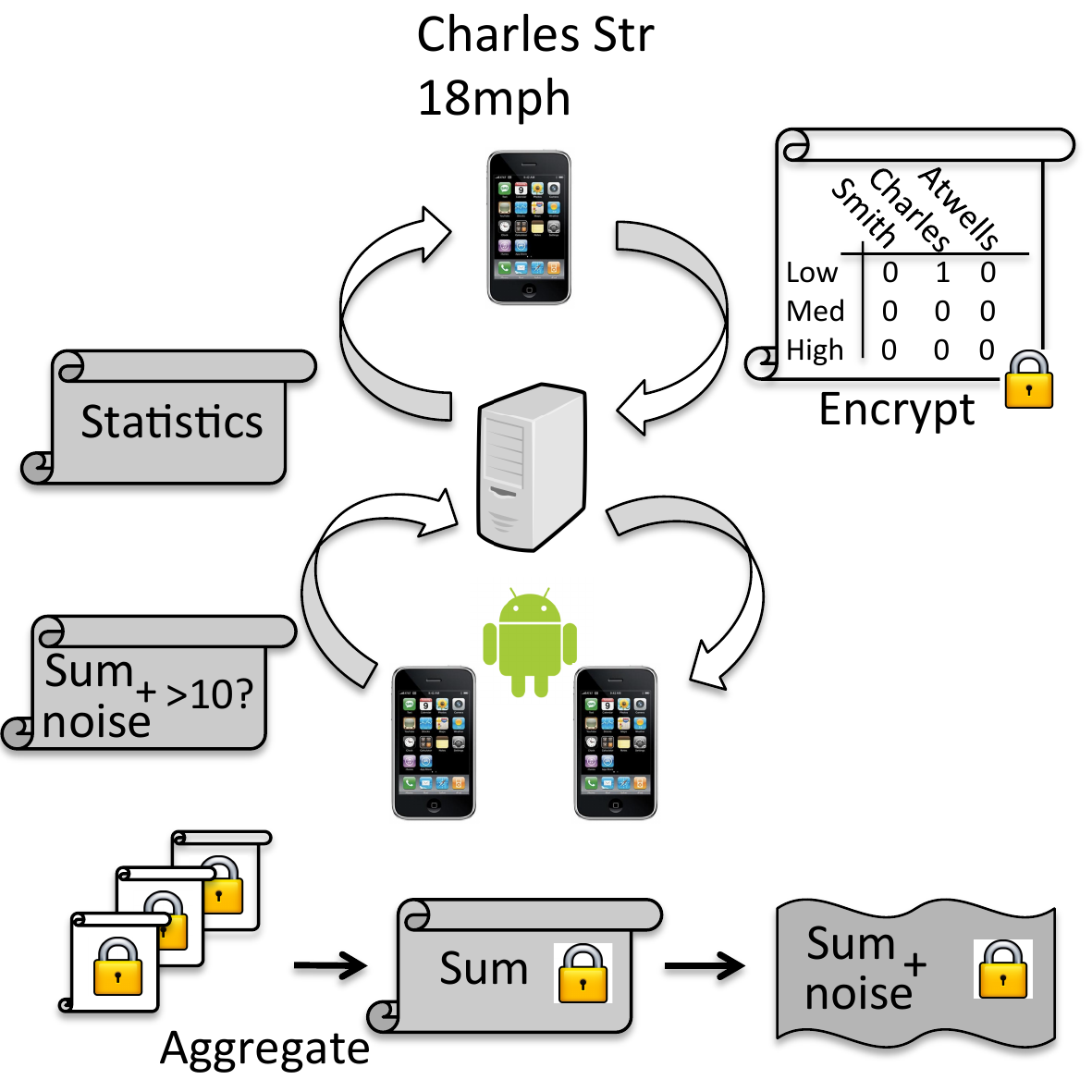}
\caption{\label{fig:datalifecycle} Data flow in Haze.}
\end{center}
\end{figure}

\fi

\ifFull

In summary, we make the following contributions:
\vspace{-10pt}

\begin{itemize}
\item We describe a protocol for real-time privacy preserving traffic data
statistics. It can report most of the information that a mobile
app for traffic and road status can, i.e., traffic flow,
presence of congestions, accidents, hazards, road work
and gas stations.
\item In contrast to previous work in this
area~\cite{priv-pbbl-11, mwpr-ppdma-13,rn-dp-10,scrcs-ppatd-11},
we model gathering of traffic data as a voting protocol which allows
us to hide the exact number of users that have made
a certain observation.
\item We design a customized differentially private mechanism
which protects an individual user vote from changing the
reported statistics by perturbing it. 
Previous work on protecting location based data either
considers a curator that has access to all the data in the clear and adds
noise to it~\cite{ahs-difftraj-12,fxs-dpmd-13},
or individual users add noise to their data before encrypting and reporting it~\cite{rn-dp-10, scrcs-ppatd-11}.
The former approach is not applicable in the distributed scenario, while
the latter may accumulate too much noise
since clients do not coordinate this phase.
Our protocol enjoys the benefits of both approaches:
the noise addition is centralized while the data comes
encrypted and has distributed origin.
\item Our protocol does not rely on a trusted third party
and pre-generated encryption keys while still offering
user privacy and protection against implausible user data.
\item We build a prototype of Haze
and test its performance on real-world dataset.
\end{itemize}

\vspace{-10pt}

The rest of this paper is organized as follows.  We describe the model
behind Haze in Section~\ref{sec:model}. The cryptographic
constructions used by Haze are overviewed in
Section~\ref{sec:crypto}. In Section~\ref{sec:protocol}, we present
the Haze protocol. In Section~\ref{sec:securityproof}, we analyze the security properties
of Haze. Work related to Haze is discussed in
Section~\ref{sec:prevwork}.
An experimental evaluation of Haze is
given in Section~\ref{sec:results}.

\else

\fi

%% file: model.tex
\section{Model}
\label{sec:model}

\ifFull
\subsection{Setting}
\fi
\label{sec:prelims}

\ifFull
\textbf{Participants}
\fi
We refer to the provider of the traffic information service
as \emph{server} and drivers that use this service, e.g.,
by downloading an app, as a set of \emph{users}.
The subset of users that participate in aggregation phase
is referred to as  \emph{authorities}.
Data aggregation is more expensive than data upload, hence,
the role of an authority is assigned at random
every time the protocol is invoked.

\ifFull
\textbf{Data}
\fi
Since the protocol is run on modern mobile services,
the application can easily determine the GPS coordinates
of a user. We assume that user speed can
be automatically determined by two timestamped GPS
coordinates. Other data such as traffic jam, accident, 
hazards, road closure, gas station, or presence of a speed
camera are manually entered by a user.
We also assume that roads and their partition into segments
by the server are publicly available and the user can determine
them from his GPS coordinates.

\ifFull
\textbf{Communication}
\fi
The users, including the authorities,
interact with each other by sending messages
to the server.
We assume the server will not drop messages since this would
result in non-accurate statistics and deteriorate her service.

\ifFull
\subsection{Aggregated Statistics}
\fi
\label{sec:stats}
Our method can report observations
about road status such as current speed,
traffic jam, construction, presence of speed cameras etc.
As noted earlier, in order to report such data,
the service provider does not need to know 
how many users have reported the presence of traffic jam,
as long as she knows that enough users
have observed it.
We model such functionality as a voting procedure:
users vote either 0 or 1 if they observed
some event or not, the votes are then tallied, and
the event is reported only if there were enough observations.
For example,
for speed reports we create several non-overlapping
speed ranges that users can vote for.
The set of all speed ranges is defined as $\mathcal{C}$
and the number of ranges in $\mathcal{C}$ depends
on the granularity of information the service provider
wants to report for the road segment $i$,
e.g., $\cats =\{(0,30), (30,60), (60,90)$
vs. $\cats =\{(0,10), (10,20), \ldots ,\allowbreak(80,90)\}$.
A user then votes for the speed range that corresponds to his travel speed.
To report the speed on segment~$i$, one can pick the range that
at least $T_i$ users have voted for.
Since several ranges in $\cats$ may have more than $T_i$ users,
$T_i$ may need to be adjusted by the service provider. 
For traffic jam reports $\cats$ is simply a singular set $\{\text{``traffic jam''}\}$.
A potential concern for the case when $|\cats| > 1$ is that a malicious
user may vote more than once. In Section~\ref{sec:upload} we show
how to prevent users from such behavior.
 
\ifFull
\subsection{Assumptions}
\fi
\label{sec:assumptions}
We assume that users that upload their
data are legitimate users and have a certified
signing public key that allows them to sign their messages
to the server. This allows anyone downloading the message
to verify that it came from a valid user and has not been tampered with.
We can relax this assumption when deploying the protocol with applications
that support social network between its users. Since in this model the users
have already established trust with their friends and family.
For example, Waze allows users to sign in with their Facebook account.

\ifFull
\subsection{Attack Model and Security Properties}
\fi
\label{sec:attackmodel}

The server is malicious and is interested in learning as much as possible
about user data and, hence, cannot be trusted.
The server may also collude with users and at most
half of the authorities. The users are not trusted with their data
reports and may try to swing
traffic status in their favor.
\ifDemo
In the view of this attack model, Haze guarantees the following properties.
\else

\label{sec:properties}

We guarantee the following properties in the
honest but curious model:
every participant behaves according to his protocol
but is curious to learn more about other users' data.
In the malicious case, the protocol can succeed and be secure 
as long as at least half of the authorities are honest. 
\fi

\emph{Server and Authority Obliviousness}~\cite{scrcs-ppatd-11}
The service provider and the authorities
should not learn anything about user data
beyond the aggregation information.

\emph{User Differential Privacy}
The user's data is protected by the differentially private mechanism
that adds noise to the statistics before it is released.
\ifFull
In particular, the output of the protocol in cases when the user
uploads his data and not are equally likely.
This property is useful when the protocol is run multiple
times and the user does not participate in all the runs.
\fi

\emph{User Accountability}
\ifFull
Since the users upload their data anonymously, they may
maliciously add misleading information such as
being on several roads at the same time, or uploading inconsistent
data (e.g., several speed measurements).
Our protocol adds a verification step
that allows to verify if the user's data is valid without learning what
it actually is.
\fi

\emph{Fault Tolerance}
Haze can report statistics as long as
at least half of the users acting as authorities
remain active during an invocation.
\ifFull
Every user that is not an authority can become inactive
as soon as he uploads the data. 
\fi
Moreover, we do not require
all legitimate users to participate in the protocol every time
it is invoked. 
\ifFull
The user may upload information whenever he
wishes to.
\fi

\ifFull
\clearpage
\fi

%% file: crypto.tex
\section{Cryptographic Primitives}
\label{sec:crypto}

Haze relies on several cryptographic primitives that
we briefly overview in this section.

\textbf{El Gamal Cryptosystem}
All user data is encrypted before it gets
sent to the authorities via the server.
We use El Gamal cryptosystem  \cite{elgamal} with 
public key $\pk$, secret key $\sk$ and 
parameters $p$ and $q$ where $q$ is
a large prime (2048 bits in our implementation) and
$p = 2qk+1$. The message space is a subgroup
$\mathbb{Z}^*_p$ of order $q$.
Encryption of message~$\msg$ from this subgroup is
denoted by $\encrypt_{\pk}(\xi, \msg)$
where $\xi$ is picked randomly.
Note that every message is encrypted together with a random value~$\xi$,
hence, encrypting the same message twice will very likely yield
a new ciphertext. For ease of notation, we write $\encrypt_{\pk}(\msg)$,
where the random nonce $\xi$ is implied.
As we will see later, no participant
can make a call to function  $\decrypt_{\sk}(\msg)$.
An important property of El Gamal cryptosystem
that we will take advantage of is additive homomorphism:
$\encrypt_{\pk}(\allowbreak \msg_1+\msg_2) = \allowbreak \encrypt_{\pk}(\msg_1)+ \encrypt_{\pk}(\msg_2)$.
(Here, operator $+$ is overloaded for simplicity since in reality 
ciphertexts are tuples of two elements, which are multiplied
correspondingly).

\textbf{Threshold Cryptosystem}
Haze protects user data from anyone participating in the protocol,
hence, there is no participant who can individually decrypt user data.
To this end, we use threshold El Gamal cryptosystem~\cite{thres} where
secret key $\sk$ is split
into shares between $n$ participants,
such that no single participant is able to determine $\sk$
nor decrypt a message encrypted under $\pk$.
Moreover, a ciphertext can be decrypted as long as at least
$n/2$ participants remain honest.

\textbf{Distributed Key Generation}
is required to generate keys for
threshold cryptosystem.
We use a distributed key generation
protocol of~\cite{gennaro_distrkeygen}, which succeeds
if at least $n/2$ participants execute the protocol
correctly. The high-level idea of this protocol
is as follows: every participant $k$ generates
$n$ secret shares $\sk_{k,k'}$ and sends one to every
other participant~$k'$. The secret share of participant $k$
is the sum of all shares received from other participants,
i.e., $\sum_{k'} \sk_{k',k}$. 
The public key is the sum of masked own shares
of every participant $k$, i.e., sum of masked $\sk_{k,k}$.
The actual protocol~\cite{gennaro_distrkeygen}
is much more involved and takes into account
participants that generate inconsistent shares.
Haze will require only a subset of users $\auths$,
aka authorities,
to participate in this protocol, and this subset
can be different every time the protocol is invoked.

\textbf{Zero Knowledge Proof}
An important feature of Haze is that all
road observations and their aggregation is performed
on encrypted data. However, malicious users may take
advantage of this, e.g., by reporting several speed
measurements or reporting traffic jam at multiple
locations.
To prevent submission of non-plausible data,
users are required to return a proof for every ciphertext
they submit. Since this proof should not reveal anything
about the actual value to the verifier, we use
zero knowledge (ZK) proof with
Fiat-Shamir heuristic to make it non-interactive~\cite{rgs-vote-97}.
In Haze, we are interested in verifying
if the ciphertext submitted by the user is an encryption
of a message $\msg$ that belongs to a set of plausible values 
(e.g., $\{0,1\}$)
without revealing which one.

\textbf{Equality Test}
To report the aggregate statistics we mentioned in
Section~\ref{sec:stats}, a set of users $\auths$
will have to determine whether the observations
of the drivers are significant to be reported or not.
The significance of the results is determined by the
number of drivers that agree on an observation on the same road.
Hence, users in the set $\auths$ will be required to perform
equality testing on encrypted values.
To determine whether $e$ is an encryption of
$\msg$, we will use the method described in~\cite{jury-hk-04}
which reduces the problem to determining if a ciphertext
is equal to 1.
The method does not require
users to decrypt $e$ but requires to generate a new
set of keys via Distributed Key Generation protocol.

\textbf{MixNet}
One of the goals of Haze is to release
useful statistics about traffic status while revealing
as little information as possible.
For example, a report of traffic jam on a particular road segment
does not require the service provider to know how many
drivers have reported it, as long as she is convinced that
a reasonable number of drivers have.
Hence, we cannot run equality tests
on the ciphertext $e$ and known threshold values.
Instead, we require the set of users, aka authorities, who run the equality test
to create a permutation of encrypted threshold values
and run equality test against this set (see below).
We employ publicly verifiable MixNet protocol to create such a permutation.
A MixNet protocol allows the authorities to encrypt and permute a set
of values, $\mathcal{S}$, in a sequence and output
the final permutation of encrypted values of $\mathcal{S}$,
such that no one, including the authorities,
can trace the values of $\mathcal{S}$ to their location in the output permutation.
The protocol proceeds as follows.
The first authority trivially encrypts
$\mathcal{S}$ as $\perm^0$, i.e., this encryption is not hiding
and everyone can verify that $\perm^0$ is an encryption of~$\mathcal{S}$.
This authority then randomly reencrypts and permutes $\perm^0$
to get a permutation $\perm^1$ and sends
it to the next authority. In general, the $k$-th authority
participating in the MixNet receives~$\perm^{k-1}$,
randomly reencrypts and permutes it, and
passes the resulting permutation, $\perm^{k}$, to the next authority.
The permutation produced by the last authority 
is the output of the MixNet. 
The process does not require
authorities to decrypt the permutation that they receive.
To make sure that every authority performs the reencryption and permutation
faithfully, we use a method by~\cite{neff-mix-01},
that allows efficient verification of the mix without
revealing the underlying permutation.

As we will see later, Haze also uses MixNets to hide the noise
added to the reported statistics.

\textbf{Inequality Test}
We perform operation $\mathsf{InequalityTest}$
to learn if an encrypted
tally $e$ of those who voted for some observation is above
a threshold $T$
or not.
We use the method
of~\cite{jury-hk-04} to determine the result
of a comparison operation without decrypting $e$.
Instead of computing
$\decrypt_\sk(\allowbreak e) \allowbreak \ge \allowbreak T$,
the authorities compute the result of
\vspace{-8pt}
\begin{equation}
\bigvee_{p=0}^{T-1} \left( \decrypt_\sk(e) \stackrel{?}{=} p \right) .
\label{eq:inequaltest}
\end{equation}

\vspace{-16pt}
The above equation can be easily computed by invoking
the equality test $T$ times.
However, we cannot run this procedure sequentially
for values $\{0,\ldots, T-1\}$, since this reveals exact count
when the number of users is below $T$.
Instead, authorities engage in the MixNet protocol
on the set $\{0,\ldots, T-1\}$.
The equality tests are then done against a
permutation of encrypted values of 0 to $T-1$.
If expression~\ref{eq:inequaltest} returns 0, then
$\decrypt_\sk(e)$ is above the threshold,
and $\mathsf{InequalityTest}$ returns $\true$.

\textbf{Differential Privacy}~\cite{d-diff-08, dmns-diff-06} is used to
preserve privacy of an individual in statistical databases.
This notion is captured by the following constraint:
Let $X$ and $X'$ be two databases which differ
only in one record, then
a mechanism $g$ is $(\epsilon, \delta)$-differentially private
if $\forall S \subseteq \mathsf{Output}(g)$
$$\Pr[g(X) \in S] \le \exp(\epsilon) \Pr[g(X') \in S] + \delta$$
In other words, an output of $g$ will be as likely
in the presence of an individual record, as in its absence.

%% file: protocol.tex
\ifFull 
\section{Haze Protocol} 
\fi
\label{sec:protocol}

\ifFull

First, we outline the Haze protocol
in the idealized case when a trusted third party (TTP) is available to assist
in the execution of the protocol by playing the role of an
intermediary between the users and the service provider.

Every time a user wishes to report a traffic observation (e.g., the
current user's speed on a given road segment), the user simply sends
it to the~TTP.  The TTP aggregates the received observations into
traffic reports (e.g., a report could state that the average speed on
a given road segment is between 40 and 50 mph) and adds noise to the
reported value to protect the privacy of the users when they are leaving
or joining the protocol.
Once the number of observations
generating a report is above a certain threshold, the TTP transmits the
report to the service provide.
Note that in this scenario, the server does not see the
association between the traffic reports forwarded by the TTP and the
users who generated them. Also, the traffic reports received by the TTP
have values that incorporate noise.

Since a protocol that employs a TTP is rather unrealistic, Haze does
not assume the existence of a~TTP. Instead, Haze simulates a TTP by
means of a distributed protocol that splits the operations of the TTP
among the server and a subset of the users, called authorities.  We
will show that we can guarantee the security properties stated in
Section~\ref{sec:properties} as long as at least half of the
authorities execute the protocol correctly.

\subsection{Outline}

\fi

The Haze protocol is invoked every time the service provider wishes to
report statistics about a certain event, e.g., average speed on a set
of roads or presence of construction work.  The protocol consists of
three phases: Setup, Data Upload, and Aggregation.

The service provider, aka the server, is involved in every phase of
the protocol while users are engaged in two of the three phases
depending on their role. In particular, a user can either contribute
to the protocol by providing his traffic observation or by assisting
in the privacy-preserving aggregation of the data provided by other
users.  We refer to the users who perform aggregations as
\emph{authorities}.

During the \emph{Setup} phase, the authorities establish encryption
keys to allow users to encrypt and hide their observations.  The
server also specifies the set of plausible candidate observations for
this event, e.g., $\cats = \{\text{``traffic jam''}\}$ or
$\mathcal{C} =\{(0,30),(30,60), (60,90)\}$.
Next, the users
participate in the \emph{Upload} phase, by sending
encrypted data to the server.  The transmitted data record consists of
the votes by the user for the candidate observations, where the vote
is either 0 or 1 depending on whether the user has actually made this
observation or not. For example, for speed ranges above
the user would vote $\{1,0,0\}$ if his speed is 18mph.
Note that authorities also can participate in this phase.
Finally, during the \emph{Result
  Computation} phase, the authorities operate on encrypted user data
to report data of the event in question.  In this phase, the
authorities tally user votes, add noise to preserve the privacy of
individual votes, and notify the server of observations that were
reported by a significant number of users.

Haze is history independent and does not rely on a fixed set of
participating users (and, hence, authorities).  Thus, the
role of a user is established during the setup phase of a single
invocation of Haze and can change the next time the protocol is run.
More importantly, the encryption keys that are created during the
setup phase are used only once.

\ifFull
In the rest of this section, we describe the three phases of the Haze
protocol.
\fi

\ifDemo
\else
The notation used is summarized in Table~\ref{tab:notation}.
\fi

\ifDemo
\else
\begin{table}[b]
\vspace{-20pt}
\caption{\label{tab:notation}%
  Notation for the Haze protocol, referring to the speed statistics example.}
\begin{center}
\begin{tabular}{r|p{6cm}}
$N$ & number of road segments \\
\hline
$M$ & number of users \\
\hline
$r(j)$, $s(j)$ & road segment and value observed by user $j$\\
\hline
$\cats$, $C=|\cats|$ & set of observation categories and its size \\
\hline
$T_{i}$ & minimum number of users for receiving speed statistics on
the $i$th road segment\\
\hline
$V_j[i][c]$ & vote by user $j$ on observation category $c$ for road
segment $i$ \\
\hline
$\pk$, $\sk$ & public key and secret key for El~Gamal cryptosystem \\
\hline
$\auths$, $A=|\auths|$ & set of authorities and its size \\
\hline
$\mathcal{N}$ &array of noises for the differentially-private mechanism
\\
\end{tabular}
\end{center}
\vspace{-12pt}
\end{table}%
\fi

\sloppy

\ifDemo
\olya{Shorten each of the following subsections.}
\fi

\subsection{Setup}
\label{sec:setup}

The server sets up the protocol by specifying a type of traffic 
for which she
wishes to collect statistics.  She then picks a set $\cats$ of
possible categories of values for this event. For
example, to collect statistics on vehicle speed,
the server can pick the following set of speed ranges:
$\cats=\{ (0,10), (10,30), (30,50), (50,\infty) \}$. 
Let $C$ denote the size of set~$\cats$.
A user reports an observation by casting a \emph{ballot}~$V$, which is a
$C$-tuple of binary votes on the
observation categories for the event.
In the above example, let $c$ denote the speed range $(30,50)$.
A user going at speed 40mph will set $V[c]=1$
and $V[d]=0$ for any other speed range $d \in \cats - {c}$.

The server also picks $N$ road segments for which she is interested in
traffic statistics.  For each road segment~$i$, she sets 
a minimum threshold~$T_i$ on the 
number of users that report an
observation for the statistics to be deemed significant.

The protocol relies on a set of users, $\auths$, called
\emph{authorities}, to perform aggregation on user data in a
privacy-preserving manner. The authorities are picked at random to
reduce the chance of selecting users that may collude with the server.
A user may flip a coin and decide if he is an authority or not.
However, he may lie about the outcome of his coin.  To make this
process publicly verifiable, we propose that users extract randomness
from a publicly verifiable random source, such as temperature at an
airport or stock-index price.

Once the set of users who serve as authorities is identified, the
authorities run the distributed key generation protocol from
Section~\ref{sec:crypto}.  By the end of this protocol, the public
key, denoted $\pk$, is announced and every authority, $k$, has a share
of the encryption secret key, denoted~$\sk_k$. By using
the threshold encryption scheme, $A/2$ authorities are enough
to decrypt the data, where $A=|\auths|$.  This approach allows Haze to be fault tolerant
in case $A/2-1$ authorities lose connectivity, and, at the same
time, allows to defend against a coalition of at most
$A/2-1$ malicious authorities who may want to try to decrypt
data of individual users.

\subsection{Data Upload}
\label{sec:upload}

\ifFull
The data upload phase is summarized in Algorithm~\ref{algo:voting}.

\textbf{Voting} 
\fi
We recall that a user casts a ballot, $V$, to indicate his
observation. The $C$ components of the ballot correspond
to the categories for the observation. Thus, the user should vote 1
for only one category and 0 for the rest.
Let $r(j)$ be the road segment that user $j$
is traveling on and let $s(j)$ be his observation (e.g., $s(j)=40$).
Instead of sending value $s(j)$,
the user casts vote 1 for the
component of the ballot associated with the category $c$ that includes
value $s(j)$, that is, the user sets $V[c]=1$ for category $c$ such
that $s(j) \in c$ (e.g., $c$ denotes the speed range $(30,50)$).
Since all the communication
between the users is done via the server, user $j$
encrypts the votes in the ballot by using the public key $\pk$, which
had been generated by the authorities in the previous step (setup).
We recall that we use a randomized encryption method. Thus, any two
votes are computationally indistinguishable.

Suppose a user is in road segment~$i$. To hide his
location, the user casts a fictitious ballots for all other
road segments in the system besides its real ballot for road
segment~$i$. Note that fictitious ballots should have all components
set to~$0$. Thus, each user submits $N$ ballots with $C$ votes each. 

\ifDemo
\else
\begin{algorithm}[hbt]
\caption{\label{algo:voting} Protocol executed
  by user $j$ in the data upload phase (Section~\ref{sec:upload})}
\begin{algorithmic}
\STATE \textbf{input:} $r(j)$: current road segment of user $j$;
    $s(j)$: value observed by user $j$ on road segment $r(j)$;.
\STATE \textbf{output:} for each road segment $i$,  ballot $V_j[i]$ cast by user
$j$ on road segment $i$, consisting of $C$ encrypted votes $V_j[i][c]$ on
every observation category~$c$; and
$\Pi_j$: an NIZK proof of the integrity of the $N$ ballots cast by user~$j$

\smallskip

\FOR{all road segments $i \in \{1,\ldots, N\}$}
	\FOR{all speed categories $c \in \cats$}
		\STATE \COMMENT{ Vote 1 for the real observation and 0 for the rest}
		\IF {$r(j) = i$ AND $s(j) \in c$}
			\STATE $V_j[i][c] \leftarrow \encrypt_\pk(1)$
		\ELSE
			\STATE $V_j[i][c] \leftarrow  \encrypt_\pk(0)$		
		\ENDIF
	\ENDFOR
         \STATE $\Pi_j[i] \leftarrow \mathsf{NIZK\_PROOF}(V[i])$
    \STATE Send ballot $V_j[i]$ to server
\ENDFOR
\STATE Send proof $\Pi_j$ of the validity of the ballots to server
\end{algorithmic}
\end{algorithm}
\fi

\ifFull
\textbf{Proving Valid Votes} Since votes are encrypted and no participant learns the
votes of other participants, a malicious user may upload invalid data.  For
example he could cast vote 2 (instead of 0 or~1), or vote for
all road segments in his neighborhood  to create the appearance of
traffic and discourage other users from
driving there.  To detect malicious behavior, we require every user to also
submit a non-interactive zero knowledge proof
(Section~\ref{sec:crypto}) of the integrity of his votes. 
\fi
 In
particular, each user  has to give a proof that in his ballot, at most one
vote is 1 and the remaining votes are~0. 
\ifFull
Recall that an NIZK proof
allows one to verify a property of a data set without learning
the data set itself.

\fi
Overall, for every road $i$, a user $j$ submits ballot $V_j[i]$ of $C$
encrypted votes. Also, user $j$ submits an NIZK proof $\Pi_j$ of the validity of the
ballots. Namely, $\Pi_j$ proves that across all ballots submitted by
user~$j$, every vote is either 0 or~1, and there is at most one vote equal
to~1.

\ifDemo
\else
\newpage

  ~ \vspace{-26pt} 
\fi

\subsection{Aggregation}
\label{sec:rescomp}
The aggregation phase begins after the server
receives and forwards to the authorities the encrypted ballots ($V_j$) and proofs of integrity ($\Pi_j$).
We assume that
the server will forward all data received since she is
interested in providing informative service to her users and stay
competitive with similar applications.
\ifFull
The aggregation phase is outlined in Algorithm~\ref{algo:auth}.
\fi

\ifDemo
\else
\begin{algorithm}[hbt!]
\caption{\label{algo:auth} Protocol executed by authority $k$ in the
  aggregation phase (Section~\ref{sec:rescomp}). }
\begin{algorithmic}

\STATE \textbf{Step~1. Verify and aggregate encrypted ballots}
\STATE  \textbf{input:} encrypted ballots and their proofs for
all the users, i.e., pairs $(V_j, \Pi_j)$ for $j = 1 \dots M$ 
\STATE   \textbf{output:} encrypted tally of the valid votes for every
road segment and category, represented by a
$C$-dimensional array $E$  of size $N$ 

  \STATE \COMMENT{Check the votes}
  \FOR {all users $j$}
      \IF {$\mathsf{verifyProof} (V_i,\Pi_i) =
        \mathsf{false}$}
        \STATE remove $j$ from the set of users 
      \ENDIF
    \ENDFOR
    \STATE \COMMENT{Aggregate votes for each road segment and category}
    \FOR{all road segments $i$ and categories $c$}
      \STATE $E[i][c] \leftarrow \sum_{j} V_j[i][c]$
    \ENDFOR
\smallskip
\STATE \textbf{Step~2. Generate noise}

\STATE \textbf{input:}  Set of noises $\noises$ 
     \COMMENT{ for differential privacy }
\STATE \textbf{output:}  for all road segments $i$,  and observation
categories, $c \in \cats$, permutation $\perm_{i,c}^{A}$ of $\noises$

\FOR{all road segments $i$ and categories $c$}
	\STATE Participate in $\mathsf{MixNet}$ protocol for $\noises$ to generate $\perm_{i,c}^{A}$
\ENDFOR
\smallskip
\STATE \textbf{Step~3. Add noise}
\STATE \COMMENT{Add noise  from the last permutation sequence to the
  encrypted tallied values}
    \FOR{all road segments $i$ and categories $c$}
      \STATE $E'[i][c] \leftarrow E[i][c] + \perm_{i,c}^{A}[1]$
    \ENDFOR

\smallskip
\STATE \textbf{Step~4. Report the statistics}
\FOR{all road segments $i \in \{1,\ldots, N\}$}
	\FOR{all observation categories $c \in \cats$}
		 \IF{$\mathsf{InequalityTest}(E'[i][c], T_i) = \true$}
		 	\STATE Report $(i,c)$  \COMMENT {$c$ was picked by at least $T_i$ users}
		\ENDIF
	\ENDFOR
\ENDFOR
\end{algorithmic}
\end{algorithm}
\fi

\ifFull
\textbf{Data Verification and Aggregation}
\fi
Every authority verifies the integrity of the users' votes
by running a non-interactive zero-knowledge verification protocol
on the encrypted ballot $V_j$ of every user $j$, using proof~$\Pi_j$.
Once all the votes are verified, every authority tallies the votes
for every road segment and category into an array~$E$.
Note that $E$ contains encrypted sum of votes since the
underlying crypto system allows addition of encrypted values.

\ifFull
\textbf{Differential Privacy}
\fi
We wish to protect every user from an adversary
who is trying to trace the user between several runs of Haze.
Since a user leaving the protocol may lead to changes
in the traffic results reported for the road he was traveling on,
and in turn reveal to the adversary the information
we are trying to hide.
We \ifFull wish to \fi protect
such attacks by adding a noise to the statistics before releasing
them to the service provider. In particular, we \ifFull wish
to \fi make the output of the protocol differentially private\ifFull :
\fi \ifFull
should be \fi bounded. 
\ifFull
 When adding noise we need
to decide on the distribution of noise, how to pick values from
this distribution, and at the same time hide which value was picked.

We define a special noise distribution as follows.
Recall that given a vector of 0/1 votes $P=[p_1, \ldots, p_M]$
we wish to compute a function $f(P,T) = 0$, if $\sum P[i] < T$,
and $f(P,T) = 1$, otherwise.
We design a customized
mechanism $g$ by picking $q$
uniformly from a set of size $1/\delta$:
$\noises = \{q~| - \lfloor {1/{2\delta}}  \rfloor < q \le  \lfloor {1}/{2\delta}  \rfloor \}$,
and
setting $g(P, T) = 0$,
if $\sum P[i] + q < T$, and $g(P, T) = 1$, otherwise.
In Section~\ref{sec:securityproof} we show that $g$
is differentially private.

To simulate the behavior of $g$,
the authorities create a permutation of the encrypted $\noises$ by invoking the MixNet protocol from Section~\ref{sec:crypto}.
The first encrypted noise of the last permutation
is then added to the final result.
As long as at least half of
the authorities execute MixNet correctly, no
authority can track noise values in $\mathcal{N}$
to values in the final permutation~$\perm^{A}$.
Each released statistics should be treated independently
when adding noise from $\noises$. Hence,
the authorities mix a total $N \times C$ noise
arrays of size~$|\mathcal{N}|$.

\textbf{Statistics Extraction}
Let $E'$ be an array of tallied encrypted values with
added noise for every road segment and observation.
To extract information from $E'$, the authorities need
to determine if the event has been observed by a significant
number of users.
In particular, for every road segment $i$
and observation category $c$, the authorities need
to check if at least $T_i$ users voted for $c$
on the $i$th segment, i.e.,
if $\decrypt_\sk(E'[i][c]) \ge T_i$.
Recall that we do not allow anyone participating in the protocol,
including authorities,
to learn the exact number of users who reported statistics for
individual road segments. Hence, the authorities are
not allowed to decrypt $E'$. Even if an authority is malicious,
he cannot decrypt $E'$ unless other $A/2-2$
authorities are malicious as well.
Moreover, as we mentioned earlier comparison over encrypted
data is not trivial and expensive.

We use the Inequality Test ( Section~\ref{sec:crypto})
for each value $E'[i][c]$ and threshold~$T_i$.
Observations for which the
test returns~$\true$ (i.e., the number of
users who voted for this observation
exceeds the threshold) are reported to the server.
\fi
Note that the users who are not authorities do not participate
during this phase, and can go offline as soon as they upload
their data. Also, users are not required to participate in every data upload
and, hence, the protocol is tolerant against any number
of users going offline.

\ifDemo
\else
\subsection{Extensions}
\label{sec:exts}

Data upload requires each user to encrypt and send
$N\times C$ messages when El~Gamal encryption is used.
One can improve the complexity
to $N+C$ if the BGN~\cite{bgn} scheme is used;
this scheme is also
additively homomorphic but allows one multiplication operation on
the ciphertexts.

Since users may be crossing multiple road segments when uploading the
data, our protocol can be easily extended to multiple-segment
voting. In this case, the ballot-validity proof should include a proof
that the user did not vote for more than the allowed number of
segments.
\fi

%% file: security.tex
\section{Security}
\label{sec:securityproof}

In this section, we show that Haze provides several formal privacy
guarantees. We begin by proving that the noise addition step yields
differential privacy.

\vspace{-16pt}
\begin{theorem}
Let $P$ be a vector of 0/1 votes $P=[p_1, \ldots, p_M ]$
and $q$ be a value picked uniformly at random
from set $\noises = \{- \lfloor {1/{2\delta}}  \rfloor+1, \ldots, \lfloor {1}/{2\delta}  \rfloor \}$.
Define mechanism $g$ as $g(P,T) = 0$, if $\sum P[i] + q < T$,
and $g(P,T) = 1$, otherwise.
We have that $g$ is $(0,\delta)$ differentially private.
\end{theorem}
\vspace{-16pt}
\begin{proof}
We show that for each $x\in \textsf{Output}(g)$, we have
\begin{eqnarray}
\Pr[g(P,T) = x] \le \exp(\epsilon) \Pr[g(P',T) = x]  + \delta
\label{eq:diffeq}:
\end{eqnarray}
where $P$ and $P'$ differ in one location, i.e.,
either $\sum P[i] = \sum P'[i] +1$ or $\sum P[i] = \sum P'[i] -1$.

We consider the case when $x=1$ (the case for $x=0$
is symmetrical) and expand Inequality~\ref{eq:diffeq}:
\begin{eqnarray}
\Pr[g(P, T) = 1]   - \Pr[g(P', T) = 1]  &=& \\
\Pr[\sum P[i]  + q \ge T]   - \Pr[\sum P'[i]  + q \ge T] 
\label{eq:ourdiff2}
\end{eqnarray}
Note that whenever $\sum P[i] = \sum P'[i] -1$,
$\Pr[\sum P[i]  + q \ge T]  \le \Pr[\sum P'[i]  + q \ge T]$
and the expression in Equation~\ref{eq:ourdiff2}
is always less than $\delta$.
Hence, we consider the case when $\sum P[i] = \sum P'[i] + 1$
in Equation~\ref{eq:ourdiff2}:
\begin{eqnarray*}
\Pr[\sum P[i]  + q \ge T]   - \Pr[\sum P[i] -1 + q \ge T]  &=& \\
 \frac{|\mathcal{N}| - T + \sum P[i] }{|\mathcal{N}|} -
 \frac{|\mathcal{N}| -T + \sum P[i] -1}{|\mathcal{N}|} &=& \delta\\
\label{eq:ourdiff3}
\end{eqnarray*}
\vspace{-20pt}
\end{proof}

Haze protects individual user privacy
even when the service provider colludes with
authorities and users, i.e., the protocol is server and authority oblivious (see
Section~\ref{sec:attackmodel}).

\vspace{-16pt}
\begin{theorem}
  Haze is server oblivious and authority oblivious if less than half of of authorities collude with the server.
\end{theorem}
\vspace{-20pt}
\begin{proof}
\emph{(Sketch)}
The privacy of user votes for an individual segment and observation
pair follows from the security of a single observation (ballot)
in the jury voting protocol in~\cite{jury-hk-04}.
We can extend the analysis to multiple segments and observations
since the only inference that the adversary can make
from seeing previous observation results, is the number of users who can potentially
vote for the rest of the ballots. This again reduces to the proof
for a single observation where the number of users participating
in voting is known,
i.e., as it is in~\cite{jury-hk-04}.
The proof depends on the correct execution of
underlying cryptographic primitives used in the protocol, i.e., Distributed Key Generation and MixNet,
which operate correctly as long as at least half of the authorities are not malicious.
If the adversary colludes with a subset of users, i.e., she knows their votes,
then she can learn if the remaining honest users voted above or below the threshold,
but cannot infer the individual votes.
\end{proof}

%% file: prevwork.tex
\section{Related Work}
\label{sec:prevwork}

\ifFull
In this section, we discuss work related to
Haze.  We begin by discussing systems with goals similar to those of
Haze. Table~\ref{tbl:prevwork} gives a comparison of Haze with
these systems.
\fi

\begin{table}[b!]
\vspace{-20pt}
\caption{\label{tbl:prevwork}
  Comparing Haze with previous work.}
\begin{center} \small
\begin{tabular}{r|c|c|c|c|c||c}
&  \begin{sideways}Carbunar \textit{et al.}~\cite{crbr-cake-13}\end{sideways} 
&  \begin{sideways}Chan \textit{et al.}~\cite{css-ppfault-12}\end{sideways} 
& \begin{sideways}PASTE~\cite{rn-dp-10}\end{sideways} 
& \begin{sideways} PrivStats~\cite{priv-pbbl-11}\end{sideways} 
& \begin{sideways}Sepia~\cite{sepia-10}\end{sideways} 
& \begin{sideways} Haze\end{sideways}  \\
\hline
Differential Privacy & &  $\checkmark$ &  $\checkmark$ & & & $\checkmark$ \\
Fault Tolerance & &  $\checkmark$ & &   $\checkmark$ & $\checkmark$ & $\checkmark$ \\
Multiple Statistics &$\checkmark$ & & &  $\checkmark$ & $\checkmark$ & $\checkmark$  \\
Distributed Key Generation & &  && &  $\checkmark$ & $\checkmark$ \\
No Trusted Third Party &  &  $\checkmark$ &   $\checkmark$ &  &  $\checkmark$ &  $\checkmark$ \\
\end{tabular}
\end{center}
\vspace{-24pt}

\end{table}%

\ifFull
\textbf{Private Aggregation of Distributed Data}
\fi
Preserving the privacy of distributed spatiotemporal data during
aggregation by an untrusted party has been studied
in~\cite{sepia-10,priv-pbbl-11,
  mwpr-ppdma-13,rn-dp-10,scrcs-ppatd-11,crbr-cake-13}.
\ifFull
PrivStats \cite{priv-pbbl-11} provides a scheme where mobile users 
report encrypted position data to location-based applications
that aggregate and report statistics on this data.
PrivStats preserves the privacy of the user's location and allows
the application to verify that users have provided
valid data. The method uses a smoothing module that adds
fictitious records and noise to the data. Also, the smoothing
module has  a secret key to the encryption scheme
and is responsible for decrypting server-aggregated statistics.
This module is installed on cars and is fully trusted.
\fi
Hence, PrivStats's trust model is very different from ours.

The system by Carbunar \textit{et al.}  \cite{crbr-cake-13} collects
aggregated location based statistics via a voting protocol. 
\ifFull
 It relies
on a trusted anonymizer between users and parties interested in
aggregated results (called venues).  Instead, Haze does not rely on a
trusted third party. Also, unlike Haze, this method does not provide
differential privacy guarantees.  The model is also different from ours
as it involves 3 parties: users, social network providers and venues.
The user receives his encryption key from the provider and sends
encrypted data to the venue, proving vote validity to him.  Haze does
not rely on receiving keys from a service provider; instead, the
authorities run a distributed key-generation protocol. Haze
also differs in the type of statistics it provides: it reports whether a vote count is higher or lower than some
threshold, but never the exact number of users.  Instead,
in~\cite{crbr-cake-13}, the number of users is revealed, since the
adversary can decrypt the tally only if exactly $k$ users have casted
their votes.
\fi

The protocols described in~\cite{mwpr-ppdma-13, rn-dp-10, scrcs-ppatd-11}
preserve distributed data aggregation by requiring each user
or node to add differentially-private noise to their data before
reporting it.
Monreale \textit{et al.}~\cite{mwpr-ppdma-13} describe a
differentially-private mechanism that allows every node
to report a trajectory of its moves, while preserving privacy of the
individual moves.
However, perturbing the reported trajectory does not hide
the link between the user and his data. Even though the server
does not know precisely when the nodes moved from one location to another,
she learns the trend of the node behavior.
On the other hand, in the framework of Rastogi and
Nath~\cite{rn-dp-10} and that of Shi \textit{et al.}~\cite{scrcs-ppatd-11},
the server never sees user data in the clear.
In their schemes, every user adds noise to his data,
encrypts it using additively homomorphic encryption
with his share of secret key,
and sends it to the server, who then aggregates received data. 
To be able to decrypt the aggregated result, in~\cite{rn-dp-10}, the server
sends the aggregated value back to every user for decryption
and then joins the partial decryptions to get the final result.
The method of~\cite{scrcs-ppatd-11} also uses an encryption scheme that allows
an aggregated value to be decrypted only if every user has contributed
his data. Chan~\textit{et al.}~\cite{css-ppfault-12} show how to make
these methods fault tolerant by requiring every user to submit their data
logarithmic number of times.
Both methods are described in the scenario where all users
are observing the data that can be aggregated into a single result,
e.g., sum over data points of the same type.
It becomes tricky to adapt these methods to traffic statistics
that we consider here. In our scenario different
users may provide observations for different roads,
and, hence, every road has its own aggregated result.
A naive extension to~\cite{rn-dp-10} and~\cite{scrcs-ppatd-11}
would allow the server to link the users and the roads they are on.
It is also worth noting that in Haze, we apply differentially private
mechanism to aggregated result, while in~\cite{rn-dp-10, scrcs-ppatd-11},
users independently apply noise before submitting their data.
Since this is a distributed mechanism, the noise variance may
reduce the utility of the aggregated result.

Sepia~\cite{sepia-10} describes a framework for private
monitoring of network traffic for multiple events.
Their approach is similar to ours
in the sense that the system also consists of input peers, who
deliver their data encrypted, and privacy peers (called authorities
in our protocol), who perform
aggregation of this data. Although the setup is similar,
the execution of data collection and aggregation is very different
from that of Haze.
In order to aggregate statistics across events, privacy peers
need to group encrypted data by their event id.
To this end, privacy peers
engage in distributed equality protocol
for every submitted encrypted user data. Once aggregated
the peers engage in distributed comparison protocol. 
Haze achieves similar functionality by using voting and mix-net
protocols. Additionally, statistics reported
by Haze are differential private.

\ifFull

Next, we overview work on the related areas of differential privacy, private
location-based queries, and jury voting.

\textbf{Differential Privacy}
Differential privacy was formulated in~\cite{dmns-diff-06}
and a survey of further results is presented in~\cite{d-diff-08}.
Differential privacy of statistical databases protects privacy of an individual
by adding noise to the released data. Noisy data adds uncertainly
on whether the data of an individual is present in the database or not.
A common scenario is to assume a presence of a curator who
has access to all the data and adds noise
before releasing it, or adds noise to query results
when operating in an interactive mode.

Differentially-private mechanisms have been
proposed to report the 
trajectory of moving objects~\cite{ahs-difftraj-12},
raw time-series traffic data~\cite{fxs-dpmd-13}
and spatial data decomposition~\cite{cpssy-diffspat-12}.
However, these methods assume
a trusted party that has access to already collected data
and is responsible for perturbing it before publishing.
This differs significantly from our work where data is distributed and
there is no trusted third party to aggregate and anonymize the data.
Differential privacy was also considered in the distributed scenario in
\cite{dkmmn-odo-06}. However, in this work all users have to engage in a distributed
noise generation protocol before releasing individual data.

\textbf{Private Location Based Queries}
Location based privacy has been considered also from the view
point of a single user who is making queries based on his location and wishes
to receive such information without revealing his location~\cite{kgmp-plbi-07,dbd-probe-10, gg-cloak-03}.
This work can be split into $k$-anonymization techniques
where user's identity cannot be distinguished from other $k$-1 users,
and cloaking techniques, where a larger region is sent during the
query instead of precise user location.

\textbf{Jury Voting} The jury voting protocol by Hevia and
Kiwi~\cite{jury-hk-04} allows participants to determine whether a vote
tally is above a threshold or not.  Haze extends the participant model
and cryptographic primitives of \cite{rgs-vote-97}
and~\cite{jury-hk-04}, to accommodate a more complicated ballot
structure and guarantee differential privacy of individual votes.

\fi

%% file: results.tex
\section{Experimental Results}
\label{sec:results}

\ifFull
We have developed a full prototype implementation of Haze
based on several primitives implemented
in the Civitas \cite{civitas} voting
system, including
distributed key generation,
El Gamal cryptosytem,
distributed exponentiation (used
in equality testing), and interface to GNU GMP (a
multiple-precision arithmetic library).
The Haze prototype includes the efficient verification of a mix-protocol
from~\cite{neff-mix-01}, an extension of the NIZK ballot validity
protocol to support voting for more than one segment, our
differentially-private algorithm, and the reporting of statistics
for multiple roads.
\else
\fi
The experiments were conducted on a 32 core 2.6GHz Opteron 6282 SE
with 64GB RAM running 64bit Debian Wheezy.
\ifFull
 For the El~Gamal
cryptosystem, we used a 2048 bit key and message space of 160 bits.
The noninteractive ZK proofs use the SHA-256 function.  The timing of the
experiments was averaged across 10 runs.
\fi
We used data from TAPAS Cologne Project~\cite{colognedata}.
This dataset contains actual GPS coordinates and speeds
of drivers in the city of Cologne, as measured during a
6am--8am time period.

The reported statistics are perturbed by noise
depending on parameter $\delta$.
In Table~\ref{tbl:delta}, we show how the value of $\delta$
influences the accuracy of the results
in terms of precision and recall measurements.
The results are taken from $1,000$ data points, during one minute,
where $4,000$ users had to pick among
$3$ speed ranges on one road segment, and the threshold on the number
of users was set to $T=11$.
As expected, the accuracy of the results degrades with lower values of $\delta$.
\ifFull
\begin{table}[b]
\vspace{-22pt}
\else
\begin{table}[tb]
\fi
 \caption{\label{tbl:delta}
    Precision and recall values for
    observations reported by Haze differentially-private mechanism
    during a period of 60 seconds.}
\vspace{-8pt}
\begin{center}
\begin{tabular}{r|c|c|c}
$\delta$ & 0.5 & 0.3 & 0.1 \\
\hline
Precision & 97\%   & 98\% &  86\% \\
Recall      & 100\% & 98\% &  88\% \\
\end{tabular}
\end{center}

\vspace{-10pt}

\caption{\label{tbl:users_vs_time}
 Total time to run Haze for $N=100$ roads,
$\cats=3$, $\delta=1/3$, $|\auths|=10$ and $T=10$
for all roads.}
\vspace{-6pt}
\begin{center}
\begin{tabular}{r|c|c|c|c}
Number of Users ($M$) & 20 & 25 & 30 & 100 \\
\hline
Total Protocol Time (secs) & 351 & 314 & 274 & 40  \\
\end{tabular}
\end{center}

\vspace{-8pt}
\end{table}%

\ifFull
We optimize the equality test in the results aggregation step
by requiring all users to participate in mixing $\{0,\ldots,T_i-1\}$
sets for tuples $(i,c)$ (Algorithm~\ref{algo:auth}). Thus, the task
is distributed over all $M$ participants
and not just the authority set.
\fi
In Table~\ref{tbl:users_vs_time}, we measure the total
time to run the protocol for different number of users
but fixed number of authorities.
The reported time is a lapse of time from the start of setup protocol
till the authorities output result for every road. Since user can upload his data in parallel
and authorities can run the MixNet, ballot verification independently,
we allowed one core per protocol participant (the time for 100 participants
is estimated from the total time of the sequential execution).
It is interesting to note that our protocol takes advantage as the number
of users grows since MixNet jobs can be distributed over a larger set.

\ifFull
\begin{figure}[b!]

\vspace{-12pt}

\begin{tabular}{l l}
  
\hspace{-6mm}
\subfigure[$M=30$, $|\auths|=10$, $|\cats|=1$]{
\includegraphics[scale=0.45]{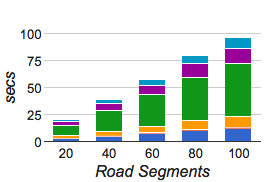}
    \label{fig:30-10-d1}
}

&
\hspace{-6mm}
\subfigure[$M=30$, $|\auths|=5$, $|\cats|=1$]{
\includegraphics[scale=0.45]{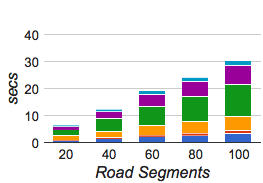}
    \label{fig:30-5-d1}
}

\\[-5pt]

\hspace{-6mm}
\subfigure[$M=30$, $|\auths|=10$, $|\cats|=3$]{
\includegraphics[scale=0.45]{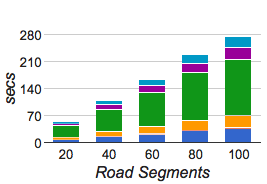}
    \label{fig:30-10-d3}
}

&

\hspace{-6mm}
\subfigure[$M=30$, $|\auths|=5$, $|\cats|=3$]{
\includegraphics[scale=0.45]{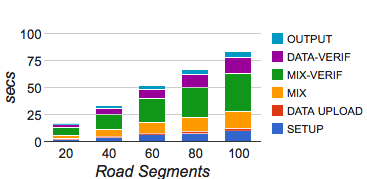}
    \label{fig:30-5-d3}
}
\end{tabular}

\fi

\ifDemo
\else
\vspace{-10pt}
\fi

\caption{\label{fig:phases}
 Breakdown of the running time of the Haze protocol by step.}

\vspace{-16pt}

\end{figure}

In Figure~\ref{fig:phases}, we show the split of the total time
across different steps in the protocol for 30 users
and 5 or 10 authorities.
As expected, the most expensive part of the protocol
is the MixNet verification.
We also vary the number of observation categories that
users can vote for: a single category in
Figures~\ref{fig:30-10-d1} and~\ref{fig:30-5-d1} (e.g., 
whether road work has been observed) and
3 categories (e.g., low/mid/high speed ranges)
in Figures~\ref{fig:30-10-d3} and~\ref{fig:30-5-d3}.
The running time of the protocol increases with the number of observation
categories.

%% file: acks.tex
\subsection*{Acknowledgments}

This research was supported in part by the National Science Foundation under grants IIS-1212508 and CNS-1228485.